\documentclass[aps,prd,showpacs,showkeys,amsmath,amssymb]{revtex4}
\usepackage{graphicx}
\usepackage{dcolumn}
\usepackage{bm}
\allowdisplaybreaks
\begin{document}

\title{Light vector meson and heavy baryon strong interaction}
\author{Peng-Zhi Huang}
\email{pzhuang@pku.edu.cn}
\author{Hua-Xing Chen}
\email{chx@water.pku.edu.cn}
\author{Shi-Lin Zhu}
\email{zhusl@pku.eud.cn}
\affiliation{Department of Physics
and State Key Laboratory of Nuclear Physics and Technology\\
Peking University, Beijing 100871, China}

\begin{abstract}

We calculate the coupling constants between the light vector mesons
and heavy baryons within the framework of the light-cone QCD sum
rule in the leading order of heavy quark effective theory. Most
resulting sum rules are stable with the variations of the Borel
parameter and the continuum threshold. The extracted couplings will
be useful in the study of the possible heavy baryon molecular
states.

\end{abstract}


\maketitle

\pagenumbering{arabic}

%%%%%%%%%%%%%%%%%%%%%%%%%%%%%%%%%%%%%%%%%%%%%%
\section{Introduction}\label{introduction}
%%%%%%%%%%%%%%%%%%%%%%%%%%%%%%%%%%%%%%%%%%%%%

Heavy quark effective theory (HQET) \cite{hqet} is a systematic
approach to study the spectra and transition amplitudes of heavy
hadrons. In HQET, the expansion is performed in terms of $1/m_Q$,
where $m_Q$ is the mass of the heavy quark involved. In the leading order of HQET,
namely $m_Q\rightarrow \infty$, heavy hadrons form a series of degenerate
doublets due to the heavy quark symmetry. The two states in a doublet
share the same quantum number $j_l$, where $j_l$ is the angular momentum of the
light components.

Light-cone QCD sum rules (LCQSR) \cite{light-cone} has been used as
a useful nonperturbative approach to extract various hadronic transition
form factors. In this framework, one considers the T-product of the two
interpolating currents sandwiched between the vacuum and an hadronic state,
instead of two vacuum states as in the conventional QCD sum rules \cite{svz}.
Now the operator product expansion is performed near the
light-cone rather than at a small distance as in
the conventional sum rules. The double Borel
transformation is always invoked to suppress the excited state and
the continuum contribution.

The $\rho$ and $\omega$ couplings between the heavy meson
doublets $H$, $S$, and $T$ were systematically studied with LCQSR in the leading
order of HQET \cite{meson}. The $\pi$ coupling constants between $\Sigma^*$, $\Sigma$, and $\Lambda$
were calculated within the same framework in Ref. \cite{zhu}.

In this work we employ LCQSR to calculate the vector meson coupling
constants between low-lying heavy baryons in the leading order of
HQET. These baryons can be grouped into two flavor multiplets
$\mathbf{6_F}$ and $\mathbf{\bar{3}_F}$, according to the flavor of
the two light quarks. On the other hand, the members in the
multiplet $\mathbf{6_F}$ are degenerate doublets in the limit
$m_Q\rightarrow \infty$. Because of the heavy quark symmetry, the vector
couplings with the same $(l,j_h)$ between two doublets are not
independent, where $l$ and $j_h$ denote the orbital and total
angular momentum of the final vector meson respectively.
Therefore the following calculations on the channels involving the $J^P=\frac{3}{2}^+$ members in $\mathbf{6_F}$
and the members in $\mathbf{\bar{3}_F}$ include all independent coupling constants under consideration.

%%%%%%%%%%%%%%%%%%%%%%%%%%%%%%%%%%%%%%%%%%%%%%%%%%%%%
\section{Sum Rules for the $\rho$ coupling constants}\label{sumrulerho}
%%%%%%%%%%%%%%%%%%%%%%%%%%%%%%%%%%%%%%%%%%%%%%%%%%%%%

According to the flavor of their two light quarks, heavy baryons can
be decomposed into two multiplets $\mathbf{6_F}$ and
$\mathbf{\bar{3}_F}$. As far as the ground states are concerned, the
total spin of the two light quarks must be 1 for
$\mathbf{6_F}$ and 0 for $\mathbf{\bar{3}_F}$ due to the symmetric
property of their colors and flavors. This leads to
$J^P=\frac{1}{2}^+/\frac{3}{2}^+$ for $\mathbf{6_F}$ and
$J^P=\frac{1}{2}^+$ for $\mathbf{\bar{3}_F}$. We use $*$ as a
superscript differentiating the $J^P=\frac{3}{2}^+$ heavy baryons
from the $J^P=\frac{1}{2}^+$ ones in $\mathbf{6_F}$.

\begin{figure}[hbt]
\begin{center}
\scalebox{0.85}{\includegraphics{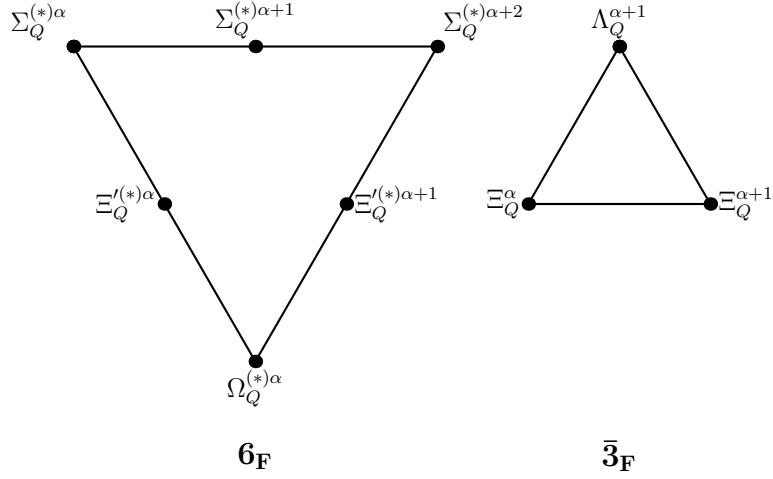}}
 \caption{The flavor $6_F$ and $\bar{3}_F$ of heavy baryons with $\alpha$, $\alpha+1$, and $\alpha+2$ denoting the
 charge of the baryon. $\alpha=-1$ or $0$ depends on the flavor of the heavy quark involved.} \label{fig:heavy_63}
\end{center}
\end{figure}

We introduce the same interpolating currents as Ref. \cite{zhu} in
our calculation for the heavy baryons with  $J^{P}=\frac{1}{2}^{+}$
in $\mathbf{\bar{3}_F}$, $J^{P}=\frac{1}{2}^{+}$ in $\mathbf{6_F}$
and $J^{P}=\frac{3}{2}^{+}$ in $\mathbf{6_F}$:
\begin{eqnarray}
\eta_{B}(x)&=&\epsilon_{abc}\Big[q_{1}^{aT}(x)C\gamma_5q_{2}^{b}(x)\Big]h_{v}^{c}(x),\\
\eta_{B'}(x)&=&\epsilon_{abc}\Big[q_{1}^{aT}(x)C\gamma_{\mu}q_{2}^{b}(x)\Big]\gamma_{t}^{\mu}\gamma_{5}h_{v}^{c}(x),\\
\eta^{\mu}_{B^*}(x)&=&\epsilon_{abc}\Big[q_{1}^{aT}(x)C\gamma_{\nu}q_{2}^{b}(x)\Big]\Big[-g_{t}^{\mu\nu}+\frac{1}{3}
\gamma^{\mu}_{t}\gamma_{t}^{\nu}\Big]h_{v}^{c}(x).
\end{eqnarray}
Here $a$, $b$, and $c$ are color indices, the subscript of $q_{i}(x)$ denotes the flavor of the quark field: $u$, $d$ or $s$.
$h_v$ is the heavy quark with velocity $v$. $T$ is the transpose matrix and $C$ is the charge conjugate matrix.
$g_t^{\mu\nu}\equiv g^{\mu\nu}-v^\mu v^\nu$, $\gamma_t^\mu\equiv \gamma^\mu-\hat{v}v^\mu$.

We denote the $\rho$ coupling constant between $\Sigma^*$ and $\Lambda$ as
$g^{p1}_{\Sigma^*\Lambda\rho}$, where the superscript $p$ and
the number $1$ indicate the orbital and total angular
momentum $(l,j_h)$ of the final $\rho$ meson respectively.
To obtain the sum rules for $g^{p1}_{\Sigma^*\Lambda\rho}$, we consider the correlation functions defined as
\begin{eqnarray}
\int d^4 xe^{-ik\cdot x}\langle \rho(q)|T\{\eta_{\Sigma^*}^\alpha (0)\bar{\eta}_\Lambda(x)\}|0\rangle
&=&\Big[g_t^{\alpha\delta}-\frac{1}{3}\gamma_t^\alpha\gamma_t^\delta\Big]\frac{1+\hat{v}}{2}
\int dt\int dx e^{-ik\cdot x}\delta(-x-vt)\nonumber\\
&&\times 4\text{Tr}[\langle\rho(q)|u(0)\bar{d}(x)|0\rangle\gamma_5 CiS^T(-x)C\gamma_\delta],
\end{eqnarray}
where $q$ is the momentum of the final $\rho$ meson, $q^2=m_\rho^2$.
In the leading order of HQET, the heavy quark propagator reads as
\begin{eqnarray}
\langle 0|T\{h_v(0)\bar{h}_v(x)\}|0\rangle=\frac{1+\hat{v}}{2}\int dt\delta^4(-x-vt).
\end{eqnarray}
$iS(-x)$ is the full light quark propagator
\begin{eqnarray}
iS(x)&\equiv&\langle 0|T\{q(x)\bar{q}(0)\}|0\rangle\nonumber\\
&=&\frac{i\hat{x}}{2\pi^2x^4}-\frac{m_q}{4\pi^2x^2}-\frac{\langle\bar{q}q\rangle}{12}+
\frac{im_q\langle\bar{q}q\rangle\hat{x}}{48}
-\frac{m^2_0\langle\bar{q}q\rangle x^2}{192}(1-\frac{im_q}{6}\hat{x})\nonumber\\
&&-\frac{ig_s}{16\pi^2}\int_0^1 du\Big\{\frac{\hat{x}}{x^2}\sigma\cdot G(ux)
-\frac{4iux_\mu\gamma_\nu}{x^2}G^{\mu\nu}(ux)
\Big\}+\cdots.
\end{eqnarray}

At the quark level, the correlation function can be calculated using
the $\rho$ meson light-cone wave functions \cite{vectormesonparameter}.
To our approximation, we need the 2- and 3-particle wave functions.

At the hadron level, the correlation function can be written as
\begin{eqnarray}
\int d^4 xe^{-ik\cdot x}\langle \rho(q)|T\{\eta_{\Sigma^*}^\alpha (0)\bar{\eta}_\Lambda(x)\}|0\rangle
=\Big[-g_t^{\alpha\delta}+\frac{1}{3}\gamma_t^\alpha\gamma_t^\delta\Big]\frac{1+\hat{v}}{2}
\epsilon^{\delta\mu\nu\rho}e^*_{t\mu}q_{t\nu}v_\rho G^{p1}_{\Sigma^*\Lambda\rho}(\omega,\omega'),
\end{eqnarray}
where $e^*_{t\mu}\equiv e^*_\mu-(e^*\cdot v)v_\mu$, $q_{t\mu}\equiv q_\mu-(q\cdot v)v_\mu$,
$\omega\equiv 2v\cdot k$, $\omega'\equiv 2v\cdot(k-q)$.
$G^{p1}_{\Sigma^*\Lambda\rho}(\omega,\omega')$ has the following pole terms
\begin{eqnarray}\label{poleterm}
\frac{4g^{p1}_{\Sigma^*\Lambda\rho}}{f_\rho}\frac{f_{\Sigma^*}f_\Lambda}{(2\bar{\Lambda}_{\Sigma^*}-\omega')(2\bar{\Lambda}_\Lambda-\omega)}
+\frac{c}{2\bar{\Lambda}_{\Sigma^*}-\omega'}+\frac{c'}{2\bar{\Lambda}_\Lambda-\omega},
\end{eqnarray}
where $\bar{\Lambda}_{\Sigma^*}\equiv m_{\Sigma^*}-m_Q$, $\bar{\Lambda}_\Lambda\equiv m_\Lambda-m_Q$.
$f_{\Sigma^*}$ etc. are the overlapping amplitudes of the interpolating currents with their corresponding states \cite{zhu}:
\begin{eqnarray} \langle
0|J_{B}|B\rangle&=&f_{B}u_{B},\\
\langle 0|J_{B'}|B'\rangle&=&f_{B'}u_{B'},\\
\langle 0|J^{\mu}_{B^{*}}|B^*\rangle&=&f_{B^*}u_{B^*} ^{\mu},
\end{eqnarray}
where  $u^{\mu}_{B^{*}}$ is the Rarita-Schwinger spinor in HQET, and we have $f_{B'}=\sqrt{3}f_{B^{*}}$
in the leading order of HQET \cite{grozin}.

We define the coupling constant $g^{p1}_{\Sigma^*\Lambda\rho}$ by the following decay amplitude
\begin{eqnarray}
\mathcal {M}(\Sigma^*\rightarrow \Lambda\rho)=\frac{g^{p1}_{\Sigma^*\Lambda\rho}}{f_\rho}\bar{u}_\Lambda u_{\Sigma^*\delta}
\epsilon^{\delta\mu\nu\rho}e^*_{t\mu}q_{t\nu}v_\rho.
\end{eqnarray}

$G^{p1}_{\Sigma^*\Lambda\rho}(\omega,\omega')$ can be expressed by the $\rho$ meson light-cone wave functions as
\begin{eqnarray}
G^{p1}_{\Sigma^*\Lambda\rho}(\omega,\omega')
&=&\frac{1}{192\pi^2}\int_0^\infty dt\int_0^1 du e^{i(1-u)\frac{wt}{2}}e^{iu\frac{w't}{2}}\nonumber\\
&&\Big[-24f_\rho^T m_\rho^2 A_T(u)\frac{1}{t}+\pi^2f_\rho m_\rho m_0^2\langle\bar{q}q\rangle g_\perp^{(a)}(u)t^3\nonumber\\
&&+16\pi^2f_\rho m_\rho\langle\bar{q}q\rangle g_\perp^{(a)}(u)t-384f_\rho^T\varphi_\perp(u)\frac{1}{t^3}\Big]\nonumber\\
&&-\frac{f_\rho^T m_\rho^2}{4\pi^2}\int_0^\infty\frac{dt}{t}\int_0^1dw
e^{i(1-\alpha_1-w\alpha_3)\frac{wt}{2}}e^{i(\alpha_1+w\alpha_3)\frac{w't}{2}}\nonumber\\
&&(\mathcal {T}_1-\mathcal {T}_2-\mathcal {T}_3+\mathcal {T}_4-\mathcal {S}-\tilde{\mathcal {S}}
+2w\mathcal {T}_3-2w\mathcal {T}_4+2w\tilde{\mathcal {S}}).
\end{eqnarray}
Here $\varphi_\perp$, $g_\perp^{(a)}$, and $A_T$ are twist-2, 3, and 4 2-particle $\rho$ meson light-cone distribution amplitudes respectively.
They are normalized to satisfy $\int_0^1\!du\, \phi(u) =1$.
$\mathcal {T}_1$, $\mathcal {T}_2$, $\mathcal {T}_3$, $\mathcal {T}_4$, $\mathcal {S}$, and $\tilde{\mathcal {S}}$
are twist-4 3-particle $\rho$ meson distribution amplitudes \cite{vectormesonparameter}.

After the Wick rotation and the double Borel transformation with $\omega$ and $\omega'$,
the single-pole terms in (\ref{poleterm}) are elminated and we arrive at
\begin{eqnarray}\label{firstsumruleintext}
g^{p1}_{\Sigma^*\Lambda\rho}f_{\Sigma^*}f_\Lambda
&=&\frac{f_\rho}{4}e^{\frac{\bar{\Lambda}_{\Sigma^*}+\bar{\Lambda}_\Lambda}{T}}\Big\{
\frac{f_\rho^T}{2\pi^2}\varphi_\perp(u_0)T^4f_3(\frac{\omega_c}{T})
-\frac{f_\rho^T m_\rho^2}{8\pi^2}A_T(u_0)T^2f_1(\frac{\omega_c}{T})\nonumber\\
&&-\frac{f_\rho m_\rho}{3}\langle\bar{q}q\rangle g_\perp^{(a)}(u_0)
+\frac{f_\rho m_\rho}{12T^2}m_0^2\langle\bar{q}q\rangle g_\perp^{(a)}(u_0)\nonumber\\
&&-\frac{f_\rho^Tm_\rho^2}{4\pi^2}T^2f_1(\frac{\omega_c}{T})
\Big[(\mathcal {T}_1-\mathcal {T}_2-\mathcal {T}_3+\mathcal {T}_4-\mathcal {S}-\tilde{\mathcal {S}})^{[0,0]}(u_0)\nonumber\\
&&+2(\mathcal {T}_3-\mathcal {T}_4+\tilde{\mathcal {S}})^{[0,1]}(u_0)\Big]
\Big\},
\end{eqnarray}
where $f_n(x)\equiv1-e^{-x}\sum_{k=0}^{n}\frac{x^k}{k!}$ is the continuum subtraction factor,
$\omega_c$ is the continnum threshold, $u_0=\frac{T_1}{T_1+T_2}$, $T\equiv\frac{T_1T_2}{T_1+T_2}$.
$T_1$ and $T_2$ are the two Borel parameters.
In the above equation we have used the following definitions
\begin{eqnarray}
\mathcal {F}^{[n]}(u_0)&\equiv&\int_0^{u_0}\cdots \int_0^{x_3}\int_0^{x_2}\mathcal {F}(x_1)dx_1dx_2\cdots dx_n,\\
\mathcal {F}^{[0,0]}(u_0)&\equiv&\int_0^{u_0}\int_{u_0-\alpha_1}^{1-\alpha_1}\frac{\mathcal {F}(\alpha_1,1-\alpha_1-\alpha_3,\alpha_3)}{\alpha_3}d\alpha_3d\alpha_1,\\
\mathcal {F}^{[0,1]}(u_0)&\equiv&\int_0^{u_0}\int_{u_0-\alpha_1}^{1-\alpha_1}\frac{(u_0-\alpha_1)\mathcal {F}(\alpha_1,1-\alpha_1-\alpha_3,\alpha_3)}{\alpha_3^2}d\alpha_3d\alpha_1,\\
\mathcal {F}^{[1,0]}(u_0)&\equiv&\int_0^{u_0}\frac{\mathcal {F}(\alpha_1,1-u_0,u_0-\alpha_1)}{u_0-\alpha_1}d\alpha_1
-\int_0^{1-u_0}\frac{\mathcal {F}(u_0,1-u_0-\alpha_3,\alpha_3)}{\alpha_3}d\alpha_3,\\
\mathcal {F}^{[1,1]}(u_0)&\equiv&\int_0^{u_0}\frac{\mathcal {F}(\alpha_1,1-u_0,u_0-\alpha_1)}{u_0-\alpha_1}d\alpha_1
-\int_0^{u_0}\int_{u_0-\alpha_1}^{1-\alpha_1}\frac{\mathcal {F}(\alpha_1,1-\alpha_1-\alpha_3,\alpha_3)}{\alpha_3^2}d\alpha_3d\alpha_1,\\
\mathcal {F}^{[-1,0]}(u_0)&\equiv&\int_0^{u_0}\int_0^{u_0-\alpha_1}
\mathcal {F}(\alpha_1,1-\alpha_1-\alpha_3,\alpha_3)d\alpha_3d\alpha_1\nonumber\\
&&+\int_0^{u_0}\int_{u_0-\alpha_1}^{1-\alpha_1}
\frac{(u_0-\alpha_1)\mathcal {F}(\alpha_1,1-\alpha_1-\alpha_3,\alpha_3)}{\alpha_3}d\alpha_3d\alpha_1,\\
\mathcal {F}^{[-1,1]}(u_0)&\equiv&\frac{1}{2}\Big\{\int_0^{u_0}\int_0^{u_0-\alpha_1}
\mathcal {F}(\alpha_1,1-\alpha_1-\alpha_3,\alpha_3)d\alpha_3d\alpha_1\nonumber\\
&&+\int_0^{u_0}\int_{u_0-\alpha_1}^{1-\alpha_1}
\frac{(u_0-\alpha_1)^2\mathcal {F}(\alpha_1,1-\alpha_1-\alpha_3,\alpha_3)}{\alpha_3^2}d\alpha_3d\alpha_1\Big\},\\
\mathcal {F}^{[-3,0]}(u_0)&\equiv&\int_0^{u_0}\int_0^{u_0-\alpha_1}\int_0^{\alpha_3}\int_0^{x_2}
\mathcal {F}(\alpha_1,1-\alpha_1-x_1,x_1)dx_1dx_2d\alpha_3d\alpha_1\nonumber\\
&&+\frac{1}{2}\int_0^{u_0}\int_0^{u_0-\alpha_1}\int_0^{\alpha_3}x\mathcal {F}(\alpha_1,1-\alpha_1-x,x)dxd\alpha_3d\alpha_1\nonumber\\
&&+\frac{1}{6}\int_0^{u_0}\int_0^{u_0-\alpha_1}\alpha_3^2\mathcal {F}(\alpha_1,1-\alpha_1-\alpha_3,\alpha_3)d\alpha_3d\alpha_1\nonumber\\
&&+\frac{1}{6}\int_0^{u_0}\int_{u_0-\alpha_1}^{1-\alpha_1}\frac{(u_0-\alpha_1)^3}{\alpha_3}\mathcal {F}(\alpha_1,1-\alpha_1-\alpha_3,\alpha_3)d\alpha_3d\alpha_1,\\
\mathcal {F}^{[-3,1]}(u_0)&\equiv&\frac{1}{2}\int_0^{u_0}\int_0^{u_0-\alpha_1}\int_0^{\alpha_3}\int_0^{x_2}
\mathcal {F}(\alpha_1,1-\alpha_1-x_1,x_1)dx_1dx_2d\alpha_3d\alpha_1\nonumber\\
&&+\frac{1}{6}\int_0^{u_0}\int_0^{u_0-\alpha_1}\int_0^{\alpha_3}x\mathcal {F}(\alpha_1,1-\alpha_1-x,x)dxd\alpha_3d\alpha_1\nonumber\\
&&+\frac{1}{24}\int_0^{u_0}\int_0^{u_0-\alpha_1}\alpha_3^2\mathcal {F}(\alpha_1,1-\alpha_1-\alpha_3,\alpha_3)d\alpha_3d\alpha_1\nonumber\\
&&+\frac{1}{24}\int_0^{u_0}\int_{u_0-\alpha_1}^{1-\alpha_1}\frac{(u_0-\alpha_1)^4}{\alpha_3^2}\mathcal {F}(\alpha_1,1-\alpha_1-\alpha_3,\alpha_3)d\alpha_3d\alpha_1
\end{eqnarray}

We have used the Borel transformation $\tilde{\mathcal {B}}_\omega^Te^{\alpha\omega}=\delta(\alpha-\frac{1}{T})$
to obtain (\ref{firstsumruleintext}).

The other $\rho$ coupling constants between $B$ and $B^*$ can be
calculated in the same way. The definitions of these coupling
constants are
\begin{eqnarray}
\mathcal {M}(\Sigma^*\rightarrow \Sigma^*\rho)
&=&\frac{g^{p0}_{\Sigma^*\Sigma^*\rho}}{f_\rho}
i\bar{u}^\alpha_{\Sigma^*}u^\xi_{\Sigma^*}g_{t\alpha\xi}\ (e^*\cdot q_t)\nonumber\\
&&+\frac{g^{p1}_{\Sigma^*\Sigma^*\rho}}{f_\rho}
i\bar{u}^\alpha_{\Sigma^*}u^\xi_{\Sigma^*}(q_{t\alpha}e^*_{t\xi}-q_{t\xi}e^*_{t\alpha})\nonumber\\
&&+\frac{g^{p2}_{\Sigma^*\Sigma^*\rho}}{f_\rho}
i\bar{u}^\alpha_{\Sigma^*}u^\xi_{\Sigma^*}\Big[q_{t\alpha}e^*_{t\xi}+q_{t\xi}e^*_{t\alpha}-\frac{2}{3}g_{t\alpha\xi}(e^*\cdot q_t)\Big]\nonumber\\
&&+\frac{g^{f2}_{\Sigma^*\Sigma^*\rho}}{f_\rho}
i\bar{u}^\alpha_{\Sigma^*}u^\xi_{\Sigma^*}(e^*\cdot
q_t)\Big[q_{t\alpha}q_{t\xi}-\frac{1}{3}g_{t\alpha\xi}q_t^2\Big],
\\
\mathcal {M}(\Xi\rightarrow \Xi\rho)
&=&\frac{g^{p0}_{\Xi\Xi\rho}}{f_\rho}i\bar{u}_\Xi u_\Xi\ (e^*\cdot
q_t).
\end{eqnarray}
The definitions of $g_{\Xi^*\Xi^*\rho}$ and $g_{\Xi^*\Xi\rho}$ are
similar to the above definitions. To derive their sum rules, we
consider the following correlators
\begin{eqnarray}
\int d^4 xe^{-ik\cdot x}\langle
\rho(q)|T\{\eta_{\Sigma^*}^\alpha(0)\bar{\eta}^{0\xi}_{\Sigma^*}(x)\}|0\rangle
&=&\Big[-g_t^{\alpha\delta}+\frac{1}{3}\gamma_t^\alpha\gamma_t^\delta\Big]\frac{1+\hat{v}}{2}
\Big[-g_t^{\xi\eta}+\frac{1}{3}\gamma_t^\eta\gamma_t^\xi\Big]
\int\!dt\int\!dx e^{-ik\cdot x}\delta(-x-vt)\nonumber\\
&&\times
4\text{Tr}\Big[\langle\rho(q)|u(0)\bar{d}(x)|0\rangle\gamma_\eta
CiS^T(-x)C\gamma_\delta\Big],
\\
\int d^4 xe^{-ik\cdot x}\langle
\rho(q)|T\{\eta_\Xi(0)\bar{\eta}^-_\Xi(x)\}|0\rangle
&=&-\frac{1+\hat{v}}{2}
\int dt\int dx e^{-ik\cdot x}\delta(-x-vt)\nonumber\\
&&\times2\text{Tr}\Big[\langle\rho(q)|u(0)\bar{d}(x)|0\rangle\gamma_5CiS_s^T(-x)C\gamma_5\Big].
\end{eqnarray}
Here $\eta^0_{\Sigma^*}(x)$ is the interpolating current for $I=0$
$\Sigma^*$ baryon.

The hadron level expression for the above two correlators reads:
\begin{eqnarray}
\int d^4 xe^{-ik\cdot x}\langle
\rho(q)|T\{\eta_{\Sigma^*}^\alpha(0)\bar{\eta}^{0\xi}_{\Sigma^*}(x)\}|0\rangle
&=&\Big[-g_t^{\alpha\delta}+\frac{1}{3}\gamma_t^\alpha\gamma_t^\delta\Big]\frac{1+\hat{v}}{2}
\Big[-g_t^{\xi\eta}+\frac{1}{3}\gamma_t^\eta\gamma_t^\xi\Big]\nonumber\\
&&\Big\{ g_{t\eta\delta}(e^*\cdot q_t)G^{p0}_{\Sigma^*\Sigma^*\rho}(\omega,\omega')\nonumber\\
&&+(q_{t\delta}e^*_{t\eta}-q_{t\eta}e^*_{t\delta})G^{p1}_{\Sigma^*\Sigma^*\rho}(\omega,\omega')\nonumber\\
&&+\Big[q_{t\delta}e^*_{t\eta}+q_{t\eta}e^*_{t\delta}-\frac{2}{3}g_{t\eta\delta}(e^*\cdot q_t)\Big]G^{p2}_{\Sigma^*\Sigma^*\rho}(\omega,\omega')\nonumber\\
&&+(e^*\cdot
q_t)\Big[q_{t\eta}q_{t\delta}-\frac{1}{3}g_{t\eta\delta}q_t^2\Big]G^{f2}_{\Sigma^*\Sigma^*\rho}(\omega,\omega')\Big\},
\\
\int d^4 xe^{-ik\cdot x}\langle
\rho(q)|T\{\eta_\Xi(0)\bar{\eta}_\Xi(x)\}|0\rangle
&=&\frac{1+\hat{v}}{2}(e^*\cdot
q_t)G^{p0}_{\Sigma\Sigma\rho}(\omega,\omega').
\end{eqnarray}
The sum rules for these coupling constants are:
\begin{eqnarray}
g^{p0}_{\Sigma^*\Sigma^*\rho}f^2_{\Sigma^*} &=&-\frac{f_\rho
m_\rho}{16\pi^2}e^{\frac{2\bar{\Lambda}_{\Sigma^*}}{T}}\Big\{
-12f_\rho\varphi_\parallel^{[1]}(u_0)T^3f_2(\frac{\omega_c}{T})
+3f_\rho
m_\rho^2\Big[A^{[1]}(u_0)+8C^{[3]}(u_0)\Big]Tf_0(\frac{\omega_c}{T})\nonumber
\\
&&+16\pi^2f_\rho^Tm_\rho\langle\bar{q}q\rangle
h_\parallel^{s[1]}(u_0)\frac{1}{T} -4\pi^2f_\rho^Tm_\rho
m_0^2\langle\bar{q}q\rangle
h_\parallel^{s[1]}(u_0)\frac{1}{T^3}\nonumber
\\
&&+2f_\rho
m_\rho^2\Big[(\Psi+2\Phi+\tilde{\Psi}+2\tilde{\Phi})^{[-1,0]}(u_0)-2(\Psi+2\Phi)^{[-1,1]}(u_0)\Big]Tf_0(\frac{\omega_c}{T})\nonumber
\\
&&-8f_\rho
m_\rho^4\Big[(\Psi-\Phi-\frac{\mathcal{V}}{2}+\tilde{\Psi}-\tilde{\Phi}-\frac{\mathcal{A}}{2})^{[-3,0]}(u_0)
-2(\Psi-\Phi-\frac{\mathcal{V}}{2})^{[-3,1]}(u_0)\Big]\frac{1}{T}\Big\}\nonumber
\\
&=&2g^{p0}_{\Xi^*\Xi^*\rho}f^2_{\Xi^*} (\Xi^*\rightarrow\Sigma^*,\
m_s\rightarrow m_q=0,\
\langle\bar{s}s\rangle\rightarrow\langle\bar{q}q\rangle),
\\
g^{p1}_{\Sigma^*\Sigma^*\rho}f^2_{\Sigma^*}
&=&-\frac{f_\rho}{32\pi^2}e^{\frac{2\bar{\Lambda}_{\Sigma^*}}{T}}\Big\{
-6f_\rho m_\rho g^{(a)}_\perp(u_0)T^3f_2(\frac{\omega_c}{T})
+16\pi^2f_\rho^T\langle\bar{q}q\rangle
\varphi_\perp(u_0)Tf_0(\frac{\omega_c}{T})\nonumber
\\
&&-4\pi^2f_\rho^T\langle\bar{q}q\rangle \Big[m^2_\rho
A_T(u_0)+m_0^2\varphi_\perp(u_0)\Big]\frac{1}{T}
+\pi^2f_\rho^Tm_\rho^2m_0^2\langle\bar{q}q\rangle
A_T(u_0)\frac{1}{T^3}\nonumber
\\
&&+3f_\rho m_\rho\Big[(\mathcal {V}+\mathcal
{A})^{[1,0]}(u_0)-2\mathcal
{A}^{[1,1]}(u_0)\Big]T^3f_2(\frac{\omega_c}{T})\nonumber
\\
&&+6f_\rho m^3_\rho\Big[(\mathcal {V}+\mathcal
{A})^{[-1,0]}(u_0)-2\mathcal
{A}^{[-1,1]}(u_0)\Big]Tf_0(\frac{\omega_c}{T})\Big\}\nonumber
\\
&=&2g^{p1}_{\Xi^*\Xi^*\rho}f^2_{\Xi^*} (\Xi^*\rightarrow\Sigma^*,\
m_s\rightarrow m_q=0,\
\langle\bar{s}s\rangle\rightarrow\langle\bar{q}q\rangle),
\\
g^{p2}_{\Sigma^*\Sigma^*\rho}f^2_{\Sigma^*} &=&-\frac{3f^2_\rho
m_\rho}{32\pi^2}e^{\frac{2\bar{\Lambda}_{\Sigma^*}}{T}}\Big\{
\Big[(\mathcal {V}+\mathcal {A})^{[1,0]}(u_0)-2\mathcal
{V}^{[1,1]}(u_0)\Big]T^3f_2(\frac{\omega_c}{T})\nonumber
\\
&&+2m^2_\rho\Big[(\mathcal {V}+\mathcal {A})^{[-1,0]}(u_0)-2\mathcal
{V}^{[-1,1]}(u_0)\Big]Tf_0(\frac{\omega_c}{T})\Big\}\nonumber
\\
&=&2g^{p2}_{\Xi^*\Xi^*\rho}f^2_{\Xi^*} (\Xi^*\rightarrow\Sigma^*,\
m_s\rightarrow m_q=0,\
\langle\bar{s}s\rangle\rightarrow\langle\bar{q}q\rangle),
\\
g^{f2}_{\Sigma^*\Sigma^*\rho}f^2_{\Sigma^*} &=&\frac{3f^2_\rho
m_\rho}{4\pi^2}e^{\frac{2\bar{\Lambda}_{\Sigma^*}}{T}}\Big\{
\Big[(\mathcal {V}+\mathcal {A})^{[-1,0]}(u_0)-2\mathcal
{V}^{[-1,1]}(u_0)\Big]Tf_0(\frac{\omega_c}{T})\nonumber
\\
&&-4m^2_\rho\Big[(\Psi-\Phi-\frac{\mathcal{V}}{2}+\tilde{\Psi}-\tilde{\Phi}-\frac{\mathcal{A}}{2})^{[-3,0]}(u_0)
-2(\Psi-\Phi-\frac{\mathcal{V}}{2})^{[-3,1]}(u_0)\Big]\frac{1}{T}\Big\}\nonumber
\\
&=&2g^{f2}_{\Xi^*\Xi^*\rho}f^2_{\Xi^*} (\Xi^*\rightarrow\Sigma^*,\
m_s\rightarrow m_q=0,\
\langle\bar{s}s\rangle\rightarrow\langle\bar{q}q\rangle),
\\
g^{p0}_{\Xi\Xi\rho}f^2_{\Xi}
&=&-\frac{1}{288\pi^2}e^{\frac{2\bar{\Lambda}_{\Xi}}{T}}f_\rho
m_\rho\Big\{ 36f_\rho
\varphi_\parallel^{[1]}(u_0)T^3f_2(\frac{\omega_c}{T}) -9f_\rho
m_\rho^2\Big[A^{[1]}(u_0)+8C^{[3]}(u_0)\Big]Tf_0(\frac{\omega_c}{T})\nonumber
\\
&&+36f_\rho^T m_\rho m_s
h_\parallel^{s[1]}(u_0)Tf_0(\frac{\omega_c}{T}) +24\pi^2f_\rho
m_s\langle\bar{s}s\rangle \varphi_\parallel^{[1]}(u_0)\frac{1}{T}
-48\pi^2f_\rho^T m_\rho\langle\bar{s}s\rangle
h_\parallel^{s[1]}(u_0)\frac{1}{T}\nonumber
\\
&&-4\pi^2f_\rho m_sm_0^2\langle\bar{s}s\rangle
\varphi_\parallel^{[1]}(u_0)\frac{1}{T^3} -6\pi^2f_\rho m_\rho^2
m_s\langle\bar{s}s\rangle
\Big[A^{[1]}(u_0)+8C^{[3]}(u_0)\Big]\frac{1}{T^3}\nonumber
\\
&&+12\pi^2f_\rho^T m_\rho m_0^2\langle\bar{s}s\rangle
h_\parallel^{s[1]}(u_0)\frac{1}{T^3} +\pi^2f_\rho
m_\rho^2m_sm_0^2\langle\bar{s}s\rangle
\Big[A^{[1]}(u_0)+8C^{[3]}(u_0)\Big]\frac{1}{T^5}\nonumber
\\
&&+18f_\rho
m_\rho^2\Big[(-\Psi-2\Phi+\tilde{\Psi}+2\tilde{\Phi})^{[-1,0]}(u_0)
+2(\Psi+2\Phi)^{[-1,1]}(u_0)\Big]Tf_0(\frac{\omega_c}{T})\nonumber
\\
&&+72f_\rho
m_\rho^4\Big[(\Psi-\Phi-\frac{\mathcal{V}}{2}-\tilde{\Psi}+\tilde{\Phi}+\frac{\mathcal{A}}{2})^{[-3,0]}(u_0)
-2(\Psi-\Phi-\frac{\mathcal{V}}{2})^{[-3,1]}(u_0)\Big]\frac{1}{T}\Big\}.
\end{eqnarray}
The sum rules for $g_{\Xi^*\Xi^*\rho}$ and ${\Xi^*\Xi\rho}$ can be
derived in a similar way and we list them below:
\begin{eqnarray}
g^{p0}_{\Xi^*\Xi^*\rho}f^2_{\Xi^*}
&=&\frac{1}{96\pi^2}e^{\frac{2\bar{\Lambda}_{\Xi^*}}{T}}f_\rho
m_\rho\Big\{ 36f_\rho
\varphi_\parallel^{[1]}(u_0)T^3f_2(\frac{\omega_c}{T}) -9f_\rho
m_\rho^2\Big[A^{[1]}(u_0)+8C^{[3]}(u_0)\Big]Tf_0(\frac{\omega_c}{T})\nonumber
\\
&&+36f_\rho^T m_\rho m_sh_\parallel^{s[1]}(u_0)Tf_0(\frac{\omega_c}{T})
+24\pi^2f_\rho m_s\langle\bar{s}s\rangle\varphi_\parallel^{[1]}(u_0)\frac{1}{T}
-48\pi^2f^T_\rho m_\rho \langle\bar{s}s\rangle h_\parallel^{s[1]}(u_0)\frac{1}{T}\nonumber
\\
&&-4\pi^2f_\rho  m_s m_0^2 \langle\bar{s}s\rangle\varphi_\parallel^{[1]}(u_0)\frac{1}{T^3}
-6\pi^2f_\rho m_\rho^2m_s\langle\bar{s}s\rangle\Big[A^{[1]}(u_0)+8C^{[3]}(u_0)\Big]\frac{1}{T^3}\nonumber
\\
&&
+12\pi^2f^T_\rho m_\rho m_0^2\langle\bar{s}s\rangle h_\parallel^{s[1]}(u_0)\frac{1}{T^3}
+\pi^2f_\rho m_\rho^2m_sm_0^2\langle\bar{s}s\rangle \Big[A^{[1]}(u_0)+8C^{[3]}(u_0)\Big]\frac{1}{T^5}\nonumber
\\
&&-6f_\rho m_\rho^2\Big[(\Psi+2\Phi+\tilde{\Psi}+2\tilde{\Phi})^{[-1,0]}(u_0)-2(\Psi+2\Phi)^{[-1,1]}(u_0)\Big]Tf_0(\frac{\omega_c}{T})\nonumber
\\
&&+24f_\rho m_\rho^4\Big[(\Psi-\Phi-\frac{\mathcal{V}}{2}+\tilde{\Psi}-\tilde{\Phi}-\frac{\mathcal{A}}{2})^{[-3,0]}(u_0)
-2(\Psi-\Phi-\frac{\mathcal{V}}{2})^{[-3,1]}(u_0)\Big]\frac{1}{T}\Big\},
\\
g^{p1}_{\Xi^*\Xi^*\rho}f^2_{\Xi^*}
&=&\frac{1}{192\pi^2}e^{\frac{2\bar{\Lambda}_{\Xi^*}}{T}}f_\rho \Big\{
36f_\rho^Tm_s\varphi_\perp(u_0)T^3f_2(\frac{\omega_c}{T})
+18f_\rho m_\rho g_\perp^{(a)}(u_0)T^3f_2(\frac{\omega_c}{T})\nonumber
\\
&&-48\pi^2f_\rho^T\langle\bar{s}s\rangle \varphi_\perp(u_0)Tf_0(\frac{\omega_c}{T})-9f_\rho^Tm_\rho^2m_sA_T(u_0)Tf_0(\frac{\omega_c}{T})
+12\pi^2f_\rho^Tm_0^2\langle\bar{s}s\rangle \varphi_\perp(u_0)\frac{1}{T}\nonumber
\\
&&+12\pi^2f_\rho^Tm_\rho^2\langle\bar{s}s\rangle A_T(u_0)\frac{1}{T}
+12\pi^2f_\rho m_\rho m_s\langle\bar{s}s\rangle g_\perp^{(a)}(u_0)\frac{1}{T}
-3\pi^2f_\rho^Tm_\rho^2m_0^2\langle\bar{s}s\rangle A_T(u_0)\frac{1}{T^3}\nonumber
\\
&&-2\pi^2f_\rho m_\rho m_s m_0^2\langle\bar{s}s\rangle g_\perp^{(a)}(u_0)\frac{1}{T^3}
-9f_\rho m_\rho\Big[(\mathcal{V}+\mathcal{A})^{[1,0]}(u_0)-2\mathcal{A}^{[1,1]}(u_0)\Big]T^3f_2(\frac{\omega_c}{T})\nonumber
\\
&&-18f_\rho m_\rho^3\Big[(\mathcal{V}+\mathcal{A})^{[-1,0]}(u_0)-2\mathcal{A}^{[-1,1]}(u_0)\Big]Tf_0(\frac{\omega_c}{T})\Big\},
\\
g^{p2}_{\Xi^*\Xi^*\rho}f^2_{\Xi^*}
&=&-\frac{3}{64\pi^2}e^{\frac{2\bar{\Lambda}_{\Xi^*}}{T}}f_\rho m_\rho\Big\{
f_\rho \Big[(\mathcal{V}+\mathcal{A})^{[1,0]}(u_0)
-2\mathcal{V}^{[1,1]}(u_0)\Big]T^3f_2(\frac{\omega_c}{T})\nonumber
\\
&&+2f_\rho m_\rho^2\Big[(\mathcal{V}+\mathcal{A})^{[-1,0]}(u_0)-2\mathcal{V}^{[-1,1]}(u_0)\Big]Tf_0(\frac{\omega_c}{T})\Big\},
\\
g^{f2}_{\Xi^*\Xi^*\rho}f^2_{\Xi^*}
&=&\frac{3}{8\pi^2}e^{\frac{2\bar{\Lambda}_{\Xi^*}}{T}}f_\rho m_\rho\Big\{
f_\rho \Big[(\mathcal{V}+\mathcal{A})^{[-1,0]}(u_0)-2\mathcal{V}^{[-1,1]}(u_0)\Big]Tf_0(\frac{\omega_c}{T})\nonumber
\\
&&-4f_\rho m_\rho^2\Big[(\Psi-\Phi-\frac{\mathcal{V}}{2}+\tilde{\Psi}-\tilde{\Phi}-\frac{\mathcal{A}}{2})^{[-3,0]}(u_0)
-2(\Psi-\Phi-\frac{\mathcal{V}}{2})^{[-3,1]}(u_0)\Big]\frac{1}{T}\Big\},
\\
g^{p1}_{\Xi^*\Xi\rho}f_{\Xi^*}f_{\Xi}
&=&\frac{i}{576\pi^2}e^{\frac{\bar{\Lambda}_{\Xi^*}+\bar{\Lambda}_{\Xi}}{T}}f_\rho \Big\{
36f_\rho^T \varphi_\perp(u_0)T^4f_3(\frac{\omega_c}{T})
-9f_\rho^Tm_\rho^2A_T(u_0)T^2f_1(\frac{\omega_c}{T})\nonumber
\\
&&+18f_\rho m_\rho m_sg_\perp^{(a)}(u_0)T^2f_1(\frac{\omega_c}{T})
+24\pi^2f_\rho^Tm_s\langle\bar{s}s\rangle \varphi_\perp(u_0)
-24\pi^2f_\rho m_\rho\langle\bar{s}s\rangle g_\perp^{(a)}(u_0)\nonumber
\\
&&-4\pi^2f_\rho^T m_sm_0^2\langle\bar{s}s\rangle \varphi_\perp(u_0)\frac{1}{T^2}
-6\pi^2f_\rho^T m_\rho^2 m_s\langle\bar{s}s\rangle A_T(u_0)\frac{1}{T^2}
+6\pi^2f_\rho m_\rho m_0^2\langle\bar{s}s\rangle g_\perp^{(a)}(u_0)\frac{1}{T^2}\nonumber
\\
&&+\pi^2f_\rho^T m_\rho^2 m_sm_0^2\langle\bar{s}s\rangle A_T(u_0)\frac{1}{T^4}
-18f_\rho^T m_\rho^2\Big[
(\mathcal{T}_1-\mathcal{T}_2-\mathcal{T}_3+\mathcal{T}_4-\mathcal{S}-\tilde{\mathcal{S}})^{[0,0]}(u_0)\nonumber
\\
&&+2(\mathcal{T}_3-\mathcal{T}_4+\tilde{\mathcal{S}})^{[0,1]}(u_0)\Big]T^2f_1(\frac{\omega_c}{T})\Big\}.
\end{eqnarray}

%%%%%%%%%%%%%%%%%%%%%%%%%%%%%%%%%%%%%%%%
\section{numerical analysis} \label{numericalanalysis}
%%%%%%%%%%%%%%%%%%%%%%%%%%%%%%%%%%%%%%%

We need the mass parameters $\bar{\Lambda}$'s and the overlapping amplitudes of these
interpolating currents $f$'s in our numerical analysis.
The values of these parameters can be extracted from the mass sum rules derived in Ref. \cite{liuxiang}.
We adopt the following values in our calculation,
noticing that the $f_{B^*}$s defined in Ref. \cite{liuxiang} is $\sqrt{3}$ times of those used in our work:

\begin{center}
\setlength\extrarowheight{8pt}
\begin{tabular}{cccccccccccc}
                        &  &$\Sigma^*$ & $\Xi^*$ & $\Omega^*$ & $\Lambda$  & $\Xi$    \\\hline
  $\bar{\Lambda}$[GeV]  &  &$1.0$      & $1.12$  & $1.25$     & $0.8$      & $1.0$    \\\hline
  $f$[GeV$^3$]          &  &$0.023$    & $0.028$ & $0.037$    & $0.018$    & $0.024$  \\
\end{tabular}
\end{center}

We take the distribution amplitudes of the $\rho$ meson
(and the $\omega$, $\phi$, and $K^*$ mesons involved in the subsequent sections) from Ref. \cite{vectormesonparameter},
there the SU(3)-breaking contributions to high-order conformal partial waves were considered in the framework of renormalon model.
Also, the parameters that appear in the distribution amplitudes of the various mesons
take the values from Ref. \cite{vectormesonparameter}.
We use the values at the scale $\mu=1\ \text{GeV}$ in our calculation
under the consideration that the heavy quark behaves almost as a spectator of the decay processes in our discussion in the leading
order of HQET.

We will work at the symmetry point, i.e., $T_1=T_2=2T$, $u_0=1/2$.
This comes from the observation that the mass difference between the baryons involved
are less than $0.45\ \text{GeV}$ in the leading
order of HQET. They are much smaller than the Borel parameter
$T_1$ , $T_2$ used below. On the other hand,
every reliable sum rule has a working window of the Borel
parameter T within which the sum rule is insensitive to the
variation of T. So it is reasonable to choose a common point
$T_1=T_2$ at the overlapping region of $T_1$ and $T_2$.
Furthermore, choosing $T_1=T_2$ will enable us to subtract the
continuum contribution cleanly while the asymmetric choice will
lead to the very difficult continuum substraction \cite{asymmetricpoint}.

\begin{figure}[hbt]
\begin{center}
\scalebox{0.85}{\includegraphics{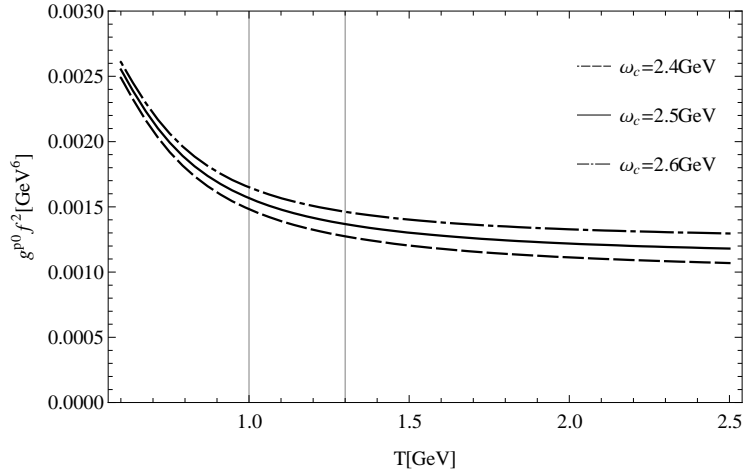}}
 \caption{The sum rule for $g_{\Sigma^*\Sigma^*\rho}^{p0}f_{\Sigma^*}^2$ with $\omega_c=2.4,2.5,2.6\ \text{GeV}$} \label{fig:P0SSRho.eps}
\end{center}
\end{figure}

\begin{figure}[hbt]
\begin{center}
\scalebox{0.85}{\includegraphics{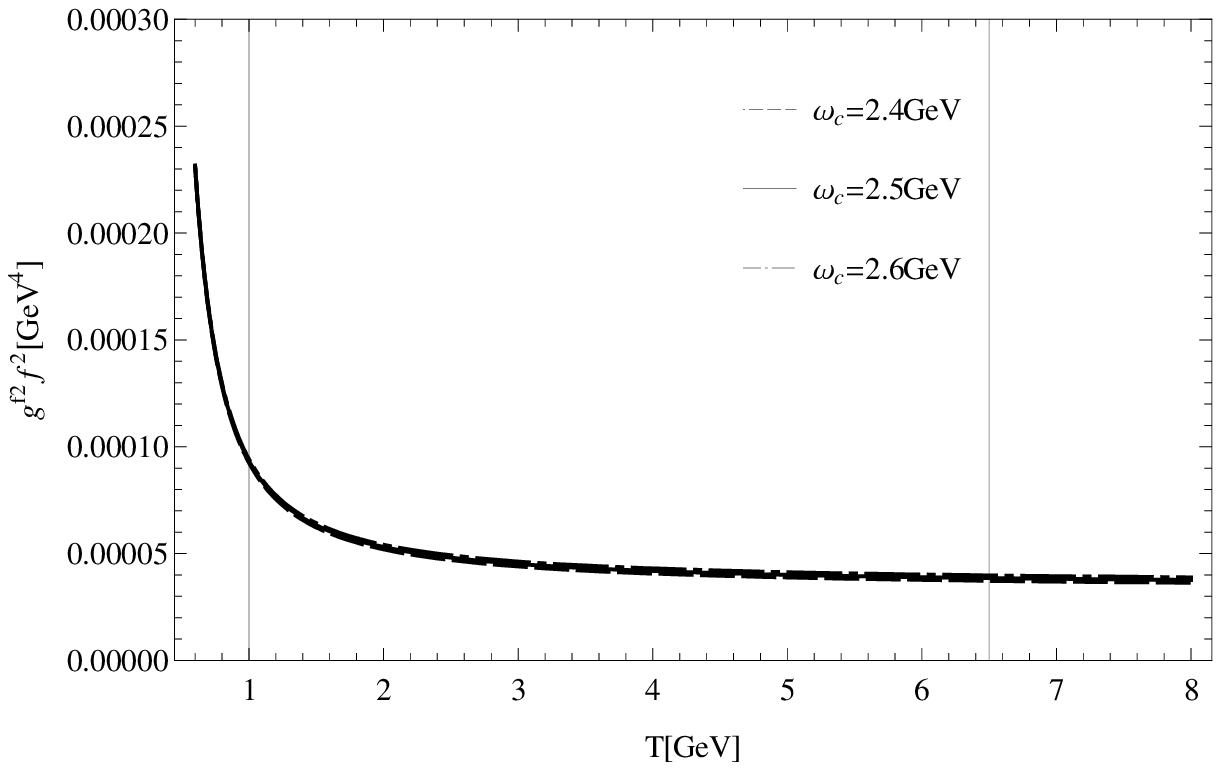}}
 \caption{The sum rule for $g_{\Sigma^*\Sigma^*\rho}^{f2}f_{\Sigma^*}^2$ with $\omega_c=2.4,2.5,2.6\ \text{GeV}$} \label{fig:F2SSRho.eps}
\end{center}
\end{figure}

From the requirement that the pole contribution is larger than
$30\%$, we get the upper bound $T_{max}$ of the Borel parameter.
In principle, every decay channel has its own $T_{max}$.
The convergence requirement of the
operator product expansion leads to the lower bound of the Borel
parameter $T_{min}=1.0\ \text{GeV}$, starting from which the stability
of the sum rule develops. We find that several sum rules have no working interval, namely $T_{min}>T_{max}$.
This may arise from the high dimension of the baryon interpolating currents.
Two typical resulting sum rules are plotted in
Fig.\ref{fig:P0SSRho.eps} and \ref{fig:F2SSRho.eps}
in the working interval $1.0\ \text{GeV}<T<1.3\ \text{GeV}$ and $1.0\ \text{GeV}<T<6.5\ \text{GeV}$,
with $\omega_c=2.4,2.5,2.6\ \text{GeV}$.

Now we can extract the numerical value of these coupling constants using the following input parameters \cite{vectormesonparameter}:
$f_\rho=0.216\ \text{GeV}$, $f_\rho^T=0.165\ \text{GeV}$, and $m_\rho=0.77\ \text{GeV}$.
The masses of $u$, $d$, and $s$ quark and their condensations are taken as:
$m_u=m_d=0$, $m_s=0.133\ \text{GeV}$, and
$\langle \bar{u}u\rangle=\langle \bar{d}d\rangle=-0.24^3\ \text{GeV}^3$, $m_0^2=0.8\ \text{GeV}^2$.

\begin{center}
\setlength\extrarowheight{8pt}
\begin{tabular}{c|ccccccccccc}
       & & $g_{\Sigma^*\Lambda\rho}$ & $g_{\Sigma^*\Sigma^*\rho}$ &
         $g_{\Xi^*\Xi^*\rho}$ & $g_{\Xi^*\Xi\rho}$  & $g_{\Xi\Xi\rho}$    \\\hline
  $p0$ & &                           &        $2.65\pm0.20$            &   $1.15\pm0.13$                  &                  &    $-1.22\pm0.17$            \\
  $p1$ & &     $\times$              &        $2.27\pm0.20$            &   $0.90\pm0.13$                  &     $\times$     &   \\
  $p2$ & &                           &        $-0.02\pm0.004$          &   $-0.005\pm0.001$               &                  &  \\
  $f2$ & &                           & $0.10\pm0.004\ \text{GeV}^{-2}$ & $0.03\pm0.001\ \text{GeV}^{-2}$  &                  &  \\
\end{tabular}
\end{center}
where ``$\times$'' indicates the non-existence of a working interval of the corresponding sum rule.
The errors come from the variations of $T$ and $\omega_c$ in the working region
and the central value corresponds to $(T_{min}+T_{max})/2$ and $\omega_c=2.5\ \text{GeV}$.

%%%%%%%%%%%%%%%%%%%%%%%%%%%%%%%%%%%%%%%%
\section{Sum Rules for the $\omega$, $K^*$, and $\phi$ coupling constants} \label{sumruleomega}
%%%%%%%%%%%%%%%%%%%%%%%%%%%%%%%%%%%%%%%

Replacing the $\rho$ meson parameters by those for the $\omega$
meson, one obtains the sum rules for the $\omega$ meson couplings
with the heavy mesons:

\begin{eqnarray}
g_{\Sigma^*\Sigma^*\omega}f_{\Sigma^*}^2
&=&\sqrt{2}g_{\Xi^*\Xi^*\rho}f_{\Xi^*}^2
(\Xi^*\rightarrow\Sigma^*,\ \rho\rightarrow\omega,\ m_s\rightarrow m_q,\ \langle\bar{s}s\rangle\rightarrow \langle\bar{q}q\rangle),
\\
g^{p0}_{\Lambda\Lambda\omega}f_{\Lambda}^2
&=&\sqrt{2}g^{p0}_{\Xi\Xi\rho}f_{\Xi}^2
(\Xi\rightarrow\Lambda,\ \rho\rightarrow\omega,\ m_s\rightarrow m_q,\ \langle\bar{s}s\rangle\rightarrow \langle\bar{q}q\rangle),
\\
g_{\Xi^*\Xi^*\omega}f_{\Xi^*}^2
&=&\frac{\sqrt{2}}{2}g_{\Xi^*\Xi^*\rho}f_{\Xi^*}^2
(\rho\rightarrow\omega),
\\
g^{p1}_{\Xi^*\Xi\omega}f_{\Xi^*}f_{\Xi}
&=&\frac{\sqrt{2}}{2}g^{p1}_{\Xi^*\Xi\rho}f_{\Xi^*}f_{\Xi}
(\rho\rightarrow\omega),
\\
g^{p0}_{\Xi\Xi\omega}f_{\Xi}^2
&=&\frac{\sqrt{2}}{2}g^{p0}_{\Xi\Xi\rho}f_{\Xi}^2
(\rho\rightarrow\omega),
\end{eqnarray}
where $g_{\Sigma^*\Sigma^*\omega} =g_{\Sigma^*\Sigma^*\omega}^{p0},\
g_{\Sigma^*\Sigma^*\omega}^{p1},\ g_{\Sigma^*\Sigma^*\omega}^{p2},\
g_{\Sigma^*\Sigma^*\omega}^{f2} $, etc.
It is straightforward to get the numerical values for various $\omega$ coupling constants from these sum rules.

\begin{center}
\setlength\extrarowheight{8pt}
\begin{tabular}{c|ccccccccccc}
       & & $g_{\Lambda\Lambda\omega}$ & $g_{\Sigma^*\Sigma^*\omega}$ & $g_{\Xi^*\Xi^*\omega}$ & $g_{\Xi^*\Xi\omega}$  & $g_{\Xi\Xi\omega}$    \\\hline
  $p0$ & &       $-1.85\pm0.06$       &        $1.51\pm0.18$         &   $0.64\pm0.13$        &                       &      $-0.87\pm0.15$            \\
  $p1$ & &                            &        $1.32\pm0.18$         &   $0.51\pm0.13$        &    $\times$           &     \\
  $p2$ & &                            &        $-0.009\pm0.002$      &   $-0.005\pm0.001$     &                       &    \\
  $f2$ & &                            &        $0.04\pm0.002\ \text{GeV}^{-2}$        &   $0.03\pm0.001\ \text{GeV}^{-2}$       &                       &    \\
\end{tabular}
\end{center}
Here we take the following values for the parameters $f_\omega$,$f_\omega^T$ \cite{ballPRD}, and $m_\omega$:
$f_\omega=0.195\ \mbox{GeV}$,$f_\omega^T=0.145\ \mbox{GeV}$, and $m_\omega=0.78\ \mbox{GeV}$.

The sum rules for the $\phi$ and $K^*$ coupling constants can be derived in a similar way:
\begin{eqnarray}
g_{\Xi^*\Xi^*\phi}f_{\Xi^*}^2
&=&g_{\Xi^*\Xi^*\rho}f_{\Xi^*}^2
(\rho\rightarrow\phi,\ m_s\rightarrow m_q=0,\ \langle\bar{s}s\rangle\rightarrow \langle\bar{q}q\rangle),
\\
g^{p1}_{\Xi^*\Xi\phi}f_{\Xi^*}f_{\Xi}
&=&-g^{p1}_{\Xi^*\Xi\rho}f_{\Xi^*}f_{\Xi}
(\rho\rightarrow\phi,\ m_s\rightarrow m_q=0,\ \langle\bar{s}s\rangle\rightarrow \langle\bar{q}q\rangle),
\\
g^{p0}_{\Xi\Xi\phi}f_{\Xi}^2
&=&g^{p0}_{\Xi\Xi\rho}f_{\Xi}^2
(\rho\rightarrow\phi,\ m_s\rightarrow m_q=0,\ \langle\bar{s}s\rangle\rightarrow \langle\bar{q}q\rangle),
\\
g^{p0}_{\Omega^*\Omega^*\phi}f_{\Omega^*}^2
&=&2g^{p0}_{\Xi^*\Xi^*\rho}f_{\Xi^*}^2
(\Xi^*\rightarrow\Omega^*,\ \rho\rightarrow\phi),
\end{eqnarray}
\begin{eqnarray}
g_{\Xi^*\Sigma^*K^*}f_{\Xi^*}f_{\Sigma^*}
&=&g_{\Xi^*\Xi^*\rho}f_{\Xi^*}^2
(\Xi^*\rightarrow\Sigma^*,\ \rho\rightarrow K^*,\ m_s\rightarrow m_q=0,\ \langle\bar{s}s\rangle\rightarrow \langle\bar{q}q\rangle),
\\
g_{\Omega^*\Xi^*K^*}f_{\Omega^*}f_{\Xi^*}
&=&2g_{\Xi^*\Xi^*\rho}f_{\Xi^*}^2
(\Xi^*\rightarrow\Omega^*,\ \rho\rightarrow K^*),
\\
g^{p0}_{\Xi\Lambda K^*}f_{\Xi}f_{\Lambda}
&=&g^{p0}_{\Xi\Xi\rho}f_{\Xi}^2
(\Xi\rightarrow\Lambda,\ \rho\rightarrow K^*,\ m_s\rightarrow m_q=0,\ \langle\bar{s}s\rangle\rightarrow \langle\bar{q}q\rangle),
\\
g^{p1}_{\Xi^*\Lambda K^*}f_{\Xi^*}f_{\Lambda}
&=&g^{p1}_{\Xi^*\Xi\rho}f_{\Xi^*}f_{\Xi}
(\Xi\rightarrow\Lambda,\ \rho\rightarrow K^*,\ m_s\rightarrow m_q=0,\ \langle\bar{s}s\rangle\rightarrow \langle\bar{q}q\rangle),
\\
g^{p1}_{\Omega^*\Xi K^*}f_{\Omega^*}f_{\Xi}
&=&2g^{p1}_{\Xi^*\Xi\rho}f_{\Xi^*}f_{\Xi}
(\Xi^*\rightarrow\Omega^*,\ \rho\rightarrow K^*),
\end{eqnarray}
and
\begin{eqnarray}
g^{p1}_{\Sigma^*\Xi K^*}f_{\Sigma^*}f_{\Xi}
&=&\frac{i}{576\pi^2}e^{\frac{\bar{\Lambda}_{\Sigma^*}+\bar{\Lambda}_{\Xi}}{T}}f_{K^*} \Big\{
36f_{K^*}^T \varphi_\perp(u_0)T^4f_3(\frac{\omega_c}{T})
-9f_{K^*}^Tm_{K^*}^2A_T(u_0)T^2f_1(\frac{\omega_c}{T})\nonumber
\\
&&-24\pi^2f_{K^*} m_{K^*}\langle\bar{q}q\rangle g_\perp^{(a)}(u_0)
+6\pi^2f_{K^*} m_{K^*} m_0^2\langle\bar{q}q\rangle g_\perp^{(a)}(u_0)\frac{1}{T^2}\nonumber
\\
&&+18f_{K^*}^T m_{K^*}^2\Big[
(\mathcal{T}_1-\mathcal{T}_2+\mathcal{T}_3-\mathcal{T}_4+\mathcal{S}-\tilde{\mathcal{S}})^{[0,0]}(u_0)\
-2(\mathcal{T}_3-\mathcal{T}_4-\tilde{\mathcal{S}})^{[0,1]}(u_0)\Big]T^2f_1(\frac{\omega_c}{T})\Big\}.
\end{eqnarray}

The input parameters of their numerical analysis are taken as
\cite{vectormesonparameter}: $f_\phi=0.215\ \text{GeV}$,
$f_\phi^T=0.186\ \text{GeV}$ , $m_\phi=1.020\ \text{GeV}$, and
$f_{K^*}=0.220\ \text{GeV}$, $f_{K^*}^T=0.185\ \text{GeV}$ ,
$m_{K^*}=0.89\ \text{GeV}$.

Also, we list the numerical results below:
\begin{center}
\setlength\extrarowheight{8pt}
\begin{tabular}{c|ccccccccccc}
       & & $g_{\Xi^*\Xi^*\phi}$ & $g_{\Xi^*\Xi\phi}$   &   $g_{\Xi\Xi\phi}$     & $g_{\Omega^*\Omega^*\phi}$   \\\hline
  $p0$ & &   $1.15\pm0.13$      &                      &   $-1.39\pm0.17$       &     $1.61\pm0.15$                  \\
  $p1$ & &   $0.89\pm0.13$      &        $\times$      &                        &     $1.31\pm0.15$                  \\
  $p2$ & &   $-0.008\pm0.001$   &                      &                        &     $-0.015\pm0.0015$               \\
  $f2$ & &   $0.025\pm0.001\ \text{GeV}^{-2}$    &                      &                        &     $0.037\pm0.0015\ \text{GeV}^{-2}$                \\
\end{tabular}
\end{center}

\begin{center}
\setlength\extrarowheight{8pt}
\begin{tabular}{c|ccccccccccc}
     && $g_{\Xi^*\Sigma^*K^*}$ & $g_{\Omega^*\Xi^*K^*}$ & $g_{\Xi\Lambda K^*}$ & $g_{\Xi^*\Lambda K^*}$ & $g_{\Omega^*\Xi K^*}$ & $g_{\Sigma^*\Xi K^*}$ \\\hline
$p0$ &&     $3.11\pm0.16$      &     $4.83\pm0.19$      &  $-4.63\pm0.70$      &                        &                       &   \\
$p1$ &&     $ 1.09\pm0.16$     &     $1.93\pm0.19$      &                      &      $\times$          &      $\times$         & $\times$  \\
$p2$ &&     $0.015\pm0.003$    &     $0.03\pm0.002$     &                      &                        &                       &   \\
$f2$ &&     $0.03\pm0.003\ \text{GeV}^{-2}$     &     $0.019\pm0.001\ \text{GeV}^{-2}$    &                      &                        &                       &   \\
\end{tabular}
\end{center}

%%%%%%%%%%%%%%%%%%%%%%%%%%%%%%%%%%%%%%%%
\section{conclusion} \label{conclusion}
%%%%%%%%%%%%%%%%%%%%%%%%%%%%%%%%%%%%%%%

We have calculated the light vector meson couplings with heavy baryons
multiplets $\mathbf{6_F}$ and $\mathbf{\bar{3}_F}$ in the leading
order of HQET, using the LCQSR approach. Most sum rules for these
coupling constants are stable with the variations of the Borel
parameter and the continuum threshold. Some possible sources of
the errors in our calculation come from the inherent inaccuracy of
LCQSR: the omission of the higher order terms in operator product expansion,
the choice of $\omega_c$, the variation of the coupling constant with the
Borel parameter $T$ in the working region and the approximation in
the light-cone distribution amplitudes of the vector meson. The
uncertainty in $f$'s and $\bar{\Lambda}$'s also leads to errors.
The 3-particle light-cone distribution amplitudes of the vector
meson are not well-known as the 2-particle ones. This may lead to
additional systematical errors in our calculation.

Recently, Belle observed a significant near-threshold enhancement
called the X(4630) in the $e^+e^-\rightarrow \Lambda_c\bar{\Lambda}_c$
exclusive cross section with initial-state radiation \cite{pakhlova}.
There have been extensive discussion of the X(4630), the interpretation
of which includes a conventional charmonium state, or a
baryon-antibaryon threshold effect, etc. The extracted coupling
constants may play an important role in the study of the interaction
between $\Lambda_c$ and $\bar{\Lambda}_c$, and therefore be helpful to clarify the
nature of the X(4630). Besides, they may be useful in the study of
the formation of possible molecular candidates composed of two
heavy baryons, such as $B_Q\bar{B}_Q$ or $B_QB_Q$.

%%%%%%%%%%%%%%%%%%%%%%%%%%%%%%%%%%%%%%%%
\section*{Acknowledgments}
%%%%%%%%%%%%%%%%%%%%%%%%%%%%%%%%%%%%%%%

This project is supported by the National Natural Science Foundation of China
under Grants No. 10625521, No. 10721063 and the Ministry of Science and
Technology of China (2009CB825200).

%---------------------------------------------------------------------------

\end{document}